# Multidimensional Stochastic Process Model and its Applications to Analysis of Longitudinal Data with Genetic Information

Ilya Zhbannikov
Biodemography of Aging Research Unit (BARU),
Social Science Research Institute (SSRI)
Duke University
Durham, NC, 27708
ilya.zhbannikov@duke.edu

Konstantin Arbeev,
Biodemography of Aging Research Unit (BARU),
Social Science Research Institute (SSRI)
Duke University
Durham, NC, 27708
konstantin.arbeev@duke.edu

Anatoliy Yashin
Biodemography of Aging Research Unit (BARU),
Social Science Research Institute (SSRI)
Duke University
Durham, NC, 27708
anatoliy.yashin@duke.edu

## ABSTRACT

Stochastic Process Model has many applications in analysis of longitudinal biodemographic data. Such data contain various physiological variables (sometimes known as covariates). It also can potentially contain genetic information available for all or a part of participants. Taking advantage from both genetic and non-genetic information can provide future insights into a broad range of processes describing aging-related changes in the organism. In this paper, we implemented a multi-dimensional Genetic Stochastic Process Model (GenSPM) in newly developed software tool, R-package `stpm`, which allows researchers performing such kind of analysis.

## 1. INTRODUCTION

Developing aging-related diseases is mediated by thousands of biological and physiological variables, which are undergo environmental and social factors, individual behavioral patterns. Various studies show involvement of genetic component in developing aging-related diseases and its potential effect on longevity [1-3]. Longitudinal study is probably the most important component of study and evaluating contribution of physiological variables to decline of the health/well-being status and a lifespan. Longitudinal data can also contain genetic information of individuals participated to the study. In such data genetic component represents genetic information in form of a genetic marker describing a particular allele. However, such longitudinal data often comes from different studies and often is incomplete, meaning that only a part of individuals are genotyped. Incompleteness in longitudinal data may arise from several factors, for example: (a) not all individuals are genotyped since some of those who initially participated in a longitudinal study have already deceased or left the study; (b) difficulty in gathering genetic information, which may arise from various reasons, for example, cost of genotyping or, perhaps, an individual refused to provide genetic samples; (c) this also can happen due to loss of some parts of experimental data during data preparation or storage.

Incomplete data often confounds the analysis leading to potential misleading results. Such cases are typical in epidemiological studies when a measure of a certain variable (covariate) is difficult to obtain for majority of participants. The common approach is to divide participants into two subgroups: the larger subgroup, which does not carry information of a particular variable, which is difficult to collect; and a smaller subgroup in which measures of such variable are presented. This is so-called two-stage analysis. Methods for analyzing such kind of data are well developed for regression models [4,5]. These methods employ information from the first subgroup (named the "first stage") and combine it with data from the second subgroup (or "second stage") in order to better estimate regression parameters. Using such methods can potentially improve precision of estimates in comparison to analysis based on the smaller group exclusively.

A common way of evaluating effects of genetic variability on health/well-being/survival condition is to estimate respective hazards (can be mortality rate) separately for carriers and non-carriers of particular allele (genotype). In absence of information of physical factors and processes affecting a hazard rate, an evaluation of a genetic effect is well developed in GWAS. Genetic data combined with longitudinal data can potentially, provide an opportunity of studying indirect genetic effects with trajectories of physiological variables, mediated by age, which, in turn, can model processes in organism not directly measured in pure longitudinal data. Therefore there is a need for special statistical methods and software tools that perform such kind of analysis.

The Stochastic Process Model (SPM) [6, 8, 9], and its extension, a genetic SPM ("GenSPM") [10], developed to deal with longitudinal data with presence of genetic information, and represents an important step toward joint analysis of longitudinal data (with corresponding genetic information) by considering together genetic- and non-genetic groups (genotyped and non-genotyped groups of participants). In this work, we (i) further extend the conception of GenSPM to a multi-dimensional case and (ii) provide a corresponding software tool: an R-package

'stpm', that implements GenSPM methodology. The stpm was verified thought extensive simulation and validation studies. This paper is also an attempt to partially close the problem of limited usage of the SPM methodology, arising from the lack of user-friendly software, documentation and examples, and promote and popularize it to academic and clinical audience.

## 2. METHODS

Originally, the Stochastic Process Model (SPM) was developed several decades ago at Duke University [6] and represents a general framework for modeling joint evolution of repeatedly measured variables (e.g., physiological or biological measures, also called covariates) and time-to-event outcomes observed in longitudinal studies. In other words, SPM links the stochastic dynamics of variables to the probabilities of end points (e.g., death or system failure). The dynamics of the stochastic variables is modeled by $N$-dimensional stochastic process (where $N$ represents a number of physiological variables used in the study) and has two components: the first component, which is related to the basic regularities of the age-related physiological changes and the second is a stochastic component which integrates the effects of external and internal perturbations of the dynamics of the physiological covariates.

In SPM methodology, the morbidity/mortality risk is presented as the quadratic hazard function, also known as U- or J- shaped hazard function [12-17], which is justified empirically based on many epidemiological observations for various biomarkers (see, e.g., [12,17]). The minimum of a hazard function, a paraboloid in the multivariable case, is a point (or domain) in the variable state space, which corresponds to the optimal system status (e.g., the "normal" health status) with the minimal hazard at a specific time (age).

In general, the SPM can be applied in the same manner as the Cox model with time-dependent covariates [7] (delete this reference). However, the advantage of the SPM methodology is that it takes into account the stochastic dynamics of variables assuming that the respective process satisfies a certain stochastic model, which better describes the reality in many applications. Finally, SPM allows projection/prediction of individual physiological trajectories, which opens possibilities for targeted research such as personalized prognoses.

GenSPM, presented in 2009 by Arbeev at al [10] (and further elaborated in 2014 "Joint analysis of longitudinal and time to event data", 2015 "Latent class and genetic stochastic processs model" (JSM) by Arbeev), further elaborates the basic stochastic process model conception by introducing a categorical variable, $Z$, which may be a specific value of a genetic marker or, in general, any categorical variable. Currently, $Z$ has two gradations: 0 or 1 in a genetic group of interest, assuming that $P(Z=1) = p$, $p \in [0, 1]$, were $p$ is the proportion of carriers and non-carriers of an allele in a population. Example of longitudinal data with genetic component $Z$ is provided in Table 1. In the previous study, the GenSPM model was verified on a single physiological variable but its behavior if more than one variable used is not known. The general concern is that using a single variable may not be enough for performing comprehensive hypotheses evaluations.

In this work we conducted corresponding simulation studies to evaluate second GenSPM dimension using simulation studies with two physiological variables by providing its validation and testing using one and two physiological covariates.

We also present a corresponding software tool, an R-package 'stpm', freely available for download from CRAN: https://cran.r-project.org/web/packages/stpm/ which is the first publicly available software, implementing the general SPM and its extension, GenSPM, allowing researchers analyzing and making predictions from longitudinal data with genetic component.

### 2.1 Description of algorithms

The block-scheme of the SPM is presented in Figure 1. In the specification of the SPM described in 2007 paper by Yashin and colleagues [8] the stochastic differential equation describing the age dynamics of a physiological variable (a dynamic component of the model) is:

$$dY(t) = a(Z, t)(Y(t) - f_1(Z, t))dt + b(Z, t)dW(t), Y(t = t_0). \quad (1)$$

Here in this equation, $Y(t)$ is a $k \times 1$ matrix, where $k$ is a number of covariates, which is a model dimension) describing the value of a physiological variable at a time (e.g. age) $t$. $f_1(Z,t)$ is a $k \times 1$ matrix that corresponds to the long-term average value of the stochastic process $Y(t)$, which describes a trajectory of individual variable influenced by different factors represented by a random Wiener process $W(t)$. The negative feedback coefficient $a(Z,t)$ ($k \times k$ matrix) characterizes the rate at which the stochastic process goes to its mean. In research on aging and well-being, $f_1(Z,t)$ represents the average allostatic trajectory and $a(t)$ in this case represents the adaptive capacity of the organism. Coefficient $b(Z,t)$ ($k \times 1$ matrix) characterizes a strength of the random disturbances from Wiener process $W(t)$. All of these parameters depend on $Z$ (a genetic marker having values 1 or 0).

The following function $\mu(t, Y(t))$ represents a hazard rate:

$$\mu(t, Y(t)) = \mu_0(t) + (Y(t) - f(Z, t))^* Q(Z, t)(Y(t) - f(Z, t)), \quad (2)$$

In this equation: $\mu_0(t)$ is the baseline hazard, which represents a risk when $Y(t)$ follows its optimal trajectory; $f(t)$ ($k \times 1$ matrix) represents the optimal trajectory that minimizes the risk and $Q(Z, t)$ ($k \times k$ matrix) represents a sensitivity of risk function to deviation from the norm. In general, model coefficients $a(Z, t)$, $f_1(Z, t)$, $Q(Z, t)$, $f(Z, t)$, $b(Z, t)$ and $\mu_0(t)$ are time(age)-dependent. For example, the coefficient a can be assumed as (i) -0.05 (a constant, time-independent, one-dimensional model) or (ii) $a(t) = a_0 + b_0 t$ (time-dependent), in which $a_0$ and $b_0$ are unknown parameters to be estimated. Presented model can handle, in theory, any number of physiological variables, however using many variables may lead to extensive usage of computational resources. Symbol '*' denotes transpose operation.

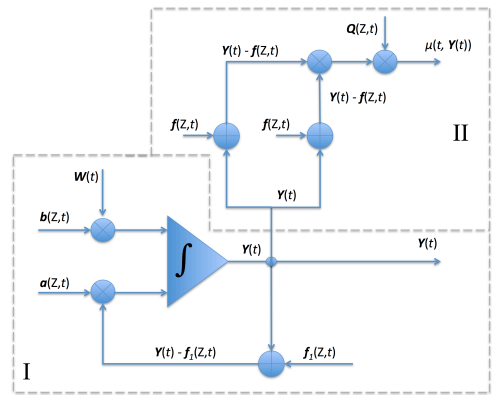

**Figure 1. Block-scheme of GenSPM. In this picture, part I is described by the equation (1) and part II is described by the equation 2 (morbidity/mortality risk).**

In order to estimate coefficients $a(Z, t)$, $f_1(Z, t)$, $Q(Z, t)$, $f(Z, t)$, $b(Z, t)$ and $\mu_0(t)$ which are the parameters of stochastic process, a method of maximizing likelihood is used (described in the next section).

| ID | Status | Age | Age.next | Z | DBP | DBP.next |
|---|---|---|---|---|---|---|
| 1 | 0 | 96.61 | 97.59 | 0 | 94.62 | 100.68 |
| 1 | 0 | 97.59 | 98.67 | 0 | 100.68 | 100.59 |
| 1 | 0 | 98.67 | 99.67 | 0 | 100.59 | 102.31 |
| 1 | 1 | 99.67 | 100.70 | 0 | 102.31 | NA |
| 2 | 0 | 64.78 | 65.78 | 1 | 81.77 | 80.62 |
| 2 | 0 | 65.78 | 66.78 | 1 | 80.62 | 70.49 |
| 2 | 0 | 66.78 | 67.68 | 1 | 70.49 | 69.20 |
| 2 | 0 | 67.68 | 68.66 | 1 | 69.20 | 67.74 |
| … | … | … | … | … | … | … |

### 2.1.1 Likelihood function for genetic group

The likelihood equations for genotyped (genetic) group were presented in [10] and is shown below:

$$L_g = p^{N_1^g(0)}(1-p)^{N_0^g(0)} \prod_{i=1}^{N_1^g(0)} L_g^i(1) \prod_{i=1}^{N_0^g(0)} L_g^i(0) \quad (3)$$

In this equation, $N_1^g(0)$ and $N_0^g(0)$ are the numbers of individuals having $Z=1$ and $0$ in the beginning of the study. The likelihood for $i$-th individual calculated as follows:

$$L_g^i(z) = \bar{\mu}^i(z,\tau_i)^{\delta_i} \exp\left\{-\int_{t_0^i}^{\tau_i} \bar{\mu}^i(z,t)\,dt\right\} \prod_{j=0}^{n_j} |\gamma^i(z,t_j^i)|^{-1/2}$$
$$\times \exp\left\{-\frac{1}{2}(y^i(t_j^i) - m^i(z,t_{j-}^i))\gamma^i(z,t_{j-}^i)^{-1}(y^i(t_j^i) - m^i(z,t_{j-}^i))^*\right\} \quad (4)$$

Then the hazard rate can be estimated with the following equation:

$$\bar{\mu}^i(z,t) = \mu_0(z,t) + (m^i(z,t) - f(z,t))^* Q(z,t)(m^i(z,t) - f(z,t))$$
$$+ Tr(Q(z,t)\gamma^i(z,t)) \quad (5)$$

Functions $m(z, t)$ and $\gamma(z, t)$ are mean and variance of the conditional distribution $P(Y(t) \le y \mid Z = z, T > t)$ that satisfy the following system of differential equations:

$$\frac{dm^i(z,t)}{dt} = a(z,t)(m^i(z,t) - f(z,t))$$
$$-2\gamma^i(z,t)Q(z,t)(m^i(z,t) - f(z,t)) \quad (6)$$

$$\frac{d\gamma^i(z,t)}{dt} = a(z,t)\gamma^i(z,t) - \gamma^i(z,t)a(z,t)^* + b(z,t)b(z,t)^*$$
$$-2\gamma^i(z,t)Q(z,t)\gamma^i(z,t)^* \quad (7)$$

dichotomous value of some genetic marker ($Z = 1$ for carriers and 0 for non-carriers). Here "ID" is a person's identification number in a database; "Status" represents death (1) or censoring (0) of an individual at age "Age.next"; "DBP" is a physiological variable, which is diastolic blood pressure in this example.

These equations are then solved at the intervals between observation times $[t_0^i, t_1^i]$, $[t_1^i, t_2^i]$,…,$[t_{n_i}^i, \tau_i]$, with initial conditions $y^i(t_0^i)$, …, $y^i(t_{n_i}^i)$ and $\gamma_{z,0}, 0,..0$ respectively. Therefore, trajectories for $m^i(z, t)$ and $\gamma^i(z, t)$ are different for each individual and so estimates of the chances of death for each individual are different; $\delta^i$ is a censoring indicator for $i$-th individual (1 – death, 0 - censored); $t_{n_i}^i$ is the last measurement of physiological variable before death/censoring at $\tau_i$ for $i$-th individual.

### 2.1.2 Likelihood function for non-genetic group

Assuming that the population of interest is heterogeneous, i.e. a mix of carriers and non-carries of particular allele or genotype, randomly selected from the data, we can write the following likelihood equation for non-genetic group [10]:

$$L_{ng} = \prod_{i=1}^{N^{ng}(0)} (pL_g^i(1) + (1-p)L_g^i(0)) \quad (8)$$

here $p$ is an expected proportion of carriers in a population; $N^{ng}(0)$ represents the number of individuals in the non-genetic group at a time (age) $t_0$. Likelihoods $L_g^i(1)$ and $L_g^i(0)$ are calculated using equation (4) above.

### 2.1.3 Joint analysis of genetic and non-genetic data

In order to combine the data from genetic and non-genetic groups, we use the following joint likelihood:

$$L = L_g L_{ng} \quad (9)$$

In this equation $L_g$ and $L_{ng}$ are likelihoods computed from genetic and non-genetic groups. This joint likelihood is then maximized used some well-known optimization method (we used Nelder-Mead [18] or COBYLA [19] optimization methods other methods are available in the R-package stpm) and thereby parameter estimates are obtained. These estimates can be used to test the hypotheses on different respective parameters for carriers and non-carriers of particular genetic marker: differences between the general model with parameters do not depend on $Z$ to the model with $Z$-dependent parameters. This, in turn, will show the presence of a genetic effect in respective component of the model.

## 2.2 R-package stpm

We developed an R-package "stpm" that comprises the SPM-methodology. The package allows for estimating several versions of SPM currently available in the literature including discrete- [20] and continuous-time multidimensional models [8] and a one-dimensional model with time-dependent parameters [9] and SPM with genetic component, described in this work. Also, the package provides routines for data preprocessing, simulation and projection of individual trajectories and hazard functions (microsimulations). The R-package stpm is available as open source software from the following link: https://cran.r-project.org/web/packages/stpm/index.html. [Add functions how to run the analysis]

The R-package `stpm` contains two following functions that basically objectify GenSPM methodology described in this article:

```
(i)     simdata_gen(…)
(ii)    spm_gen(…)
```

The `simdata_gen(…)` is used for data simulation which may be useful in case of data absence and the `spm_gen(…)` provides estimation of the parameters of the model described above. In the example below we show how to work with this function:

```
library(stpm)
#Data simulation:
data <- simdata_gen(N=100, mode='genetic')
head(data)
#Estimation of parameters:
pars <- spm_gen(gendat=data)
pars
```

Here we simulated a dataset containing $N$=100 genotyped individuals (using parameter `mode='genetic'`) and then estimated all parameters for the one-dimensional model.

In the next example we show joint analysis of two datasets: first dataset with genetic and second dataset with non-genetic component (on simulated data, as in previous example):

```
library(stpm)
# Data simulation for genetic and non-genetic group:
data.genetic <- simdata_gen(N=100, mode='genetic')
data.nongenetic <- simdata_gen(N=500, mode='nongenetic')
#Estimation of parameters:
pars <- spm_gen(gendat=data.genetic, nongendat = data.nongenetic, mode='combined')
pars
```

Here we use `mode='genetic'`, which indicates that two datasets, one from genetic (or genotyped) group and second from non-genetic (not-genotyped) group.

## 2.3 Verification strategies

In order to verify the conception presented, and test the corresponding R-package 'stpm', we conducted simulation studies. Using the R-package stpm we performed simulations and then model parameter estimations for genetic groups and then we estimated parameters from combined data (genetic and non-genetic groups). Specifically, we simulated the following datasets containing one and two variables: (i) 100 datasets for genetic group contained $N$=100 individuals; 100 datasets for genetic group of 1,000 of individuals; and 100 datasets for non-genetic group of 5,000 of individuals, with $p$ = 0.25 in all cases (Test 1); (ii) the same strategy but we used two physiological variables instead of one (Test 2). Table 2 presents the summary of these datasets. Initial values of $Y(t)$ are assumed normally distributed: $N(f_1(t_0), \sigma_0^2)$, where $f_1(t_0)$ is the value of $f_1$ in time $t$=$t_0$ ($t_0 \in$[30, 60], uniformly distributed).

**Table 2. Description of the tests:** $N$g1 **and** $N$g2 **– numbers of individuals in the genetic group (two separate genetic groups simulated),** $N$ng **– number of individuals in the non-genetic group. There were 100 datasets simulated for each group.**

| Test # | # of covariates | $N_{g1}$ | $N_{g2}$ | $N_{ng}$ | $N_{g1}$+$N_{ng}$ |
|---|---|---|---|---|---|
| 1 | 1 | 100 | 1,000 | 5,000 | 100 + 5,000 |
| 2 | 2 | 100 | 1,000 | 5,000 | 100 + 5,000 |

## 3. RESULTS
### 3.1 Test 1

Table 3 contains results for Test 1. In this test we performed evaluation of model behavior for one physiological variable. For genetic group, parameter estimates, such as $p$, Q, f1 are getting close to their true values used in simulation as number of individuals in the cohort increases. Estimates obtained from combined data: genetic (100 of individuals) + non-genetic groups (5,000 of individuals) are very close to those parameters used in simulation.

### 3.2 Test 2

Table 4 shows results for Test 2 (two physiological variables used). Even with small number of individuals (100), parameter estimates are close to their true values. As in previous case, parameter estimates are close to the parameters used in data simulations.

## 4. DISCUSSION

Longitudinal data contain measurements of various physiological variables. Such data can also contain genetic information of subjects. Often only genetic information from such data is selected for study of specific problem, leaving untouched the rest of the data. Such analysis on abridged data can lead to misleading and unpredicted results therefore it must be extended to use data from a non-genetic subgroup as well. Combining results from different studies is another example of such problem.

From the other side, analysis only of genetic component presented in longitudinal data may often not be enough since it does not offer an entire picture on aging-related changes in humans providing only an overview of genetic influence on such mechanisms. Stochastic Process Model and its extension, Genetic SPM ('GenSPM') allow researchers to utilize the entire potential of longitudinal data by evaluating dynamic mechanisms of changing physiological variables with time (age), allowing studying differences in genotype-specific hazards. Applying the Stochastic Process Model to analysis of longitudinal data uncovers influences of hidden components (adaptive capacity, allostatic load, resistance to stresses, physiological norm) of aging-related changes, which play important role in aging-related processes but cannot be measured directly with common statistical methods. This brings researchers to a new way of analyzing longitudinal data, however there are several concerns that should be taken into account before conducting such data analysis.

The first concern is an assumption that the genetic group is a random sample from the entire cohort under study. In reality this may not be true because, for example, only people with some particular disease were genotyped or some of them refused participating to genotyping or already deceased. The methods that take this into account should be elaborated.

Second concern is an assumption of specific forms of hazard risk function. In our approach we assume that the hazard rate (incidence rate related to changing physiological variable with age) has the form of U- or J- shape, which is biologically justified by empirical observations. In reality, the true form of such function is not known and since it is impossible to estimate it from the data, incorrectly assumed hazard may introduce additional bias. Additional investigations are needed in order to evaluate effects of different forms of hazard functions on results.

We performed our studies and proposed a software tool for random variable $Z$ with two gradations: 0 and 1. However, the

GenSPM can be easily extended to three values: 0, 1 and 2 corresponding to genotypes "aa", "Aa" and "AA".

In this work we presented a first software implementation a multi-dimensional genetic SPM, 'GenSPM' and corresponding software tool (an R-package 'stpm'). Future work includes further improvements of the model, e.g. introducing three genetic markers and testing on higher number of variables. Clinical and academic researchers will benefit from using presented model and software.

## 5. ACKNOWLEGEMENTS

This work was supported by the National Institute on Aging of the National Institutes of Health (NIA/NIH) under Award Numbers P01AG043352, R01AG046860, and P30AG034424. The content is solely the responsibility of the authors and does not necessarily represent the official views of the NIA/NIH.

Table 3. Results for one-dimensional simulation (one physiological variable), for two genetic and joint genetic and non-genetic groups. Values in brackets contain 95% confidence interval, lower and upper bound. True parameter values are those used in simulation. Here 'H' and 'L' represents parameters when $Z = 1$ (H) and 0 (L).

| N | aH | aL | f1H | f1L | QH x $10^{-4}$ | QL x $10^{-4}$ | fH | fL | bH | bL | $\mu_0$H x $10^{-2}$ | $\mu_0$L x $10^{-3}$ | p |
|---|---|---|---|---|---|---|---|---|---|---|---|---|---|
| 100 (genetic) | -0.202 [-0.24, -0.169] | -0.201 [-0.221, -0.18] | 40 [38.9, 41.1] | 45 [44.6, 45.5] | 0.507 [0.311, 0.679] | 0.304 [0.108, 0.491] | 40.2 [35.4, 46.2] | 49.9 [46.7, 54.6] | 5 [4.82, 5.18] | 4 [3.92, 4.07] | 0.221 [0.101, 0.38] | 0.316 [0.108, 0.494] | 0.293 [0.191, 0.399] |
| 1,000 (genetic) | -0.198 [-0.209, -0.187] | -0.2 [-0.206, -0.193] | 40 [39.6, 40.4] | 45 [44.9, 45.2] | 0.529 [0.421, 0.667] | 0.327 [0.229, 0.427] | 40.1 [38.2, 42.5] | 49.6 [48.2, 51.5] | 5 [4.94, 5.06] | 4 [3.98, 4.03] | 0.227 [0.113, 0.332] | 0.325 [0.151, 0.491] | 0.261 [0.159, 0.344] |
| 100 + 5,000 (combined) | -0.199 [-0.205, -0.19] | -0.199 [-0.203, -0.196] | 40.1 [39.7, 40.3] | 45 [44.9, 45.1] | 0.533 [0.427, 0.635] | 0.322 [0.226, 0.429] | 40.7 [39.2, 42.2] | 49.6 [48.3, 51] | 4.99 [4.94, 5.04] | 4.01 [3.99, 4.02] | 0.242 [0.144, 0.344] | 0.351 [0.173, 0.457] | 0.252 [0.228, 0.276] |
| **True** | -0.2 | -0.2 | 40 | 45 | 0.5 | 0.3 | 40 | 50 | 5 | 4 | 0.25 | 0.33 | 0.25 |

Table 4. Results for two-dimensional simulation (two physiological variables), for two genetic and joint genetic and non-genetic groups. Values in brackets contain confidence 95% interval, lower and upper bound. True parameter values are those used in simulation. For matrices *a* and *Q* only components on the main diagonal are shown. 'H' and 'L' represents parameters when $Z = 1$ (H) and 0 (L).

| N | $a_{11}$H | $a_{11}$L | $f1_1$H | $f1_1$L | $Q_{11}$H x $10^{-4}$ | $Q_{11}$L x $10^{-4}$ | $f_1$H | $b_1$H | $b_1$L | $\mu_0$H x $10^{-2}$ | p |
|---|---|---|---|---|---|---|---|---|---|---|---|
| 100 (genetic) | -0.186 [-0.241, -0.151] | -0.192 [-0.232, -0.156] | 41.7 [36.8, 46.2] | 47.8 [41.3, 56] | 0.55 [0.466, 0.65] | 0.343 [0.262, 0.439] | 40.9 [35.9, 45] | 4.98 [4.48, 5.49] | 3.97 [3.44, 4.47] | 0.27 [0.223, 0.337] | 0.26 [0.206, 0.307] |
| 1,000 (genetic) | -0.187 [-0.239, -0.153] | -0.183 [-0.228, -0.153] | 42 [36.2, 47.4] | 47.3 [41.4, 54.4] | 0.551 [0.467, 0.649] | 0.337 [0.271, 0.433] | 40.9 [37, 44.6] | 4.97 [4.51, 5.44] | 3.95 [3.39, 4.42] | 0.259 [0.214, 0.302] | 0.258 [0.216, 0.301] |
| 100 + 5,000 (combined) | -0.183 [-0.235, -0.151] | -0.19 [-0.235, -0.157] | 43.4 [38.6, 49] | 45.3 [40.3, 49.7] | 0.562 [0.475, 0.661] | 0.353 [0.261, 0.459] | 43.3 [36.9, 48.8] | 5 [4.52, 5.47] | 4.03 [3.44, 4.57] | 0.287 [0.232, 0.35] | 0.265 [0.208, 0.31] |
| True | -0.2 | -0.2 | 40 | 45 | 5e-05 | 3e-05 | 40 | 5 | 4 | 0.25 | 0.25 |

| N | $a_{22}$H | $a_{22}$L | $f1_2$H | $f1_2$L | $Q_{22}$H x $10^{-4}$ | $Q_{22}$L x $10^{-4}$ | $f_2$H | $b_2$H | $b_2$L | $\mu_0$H x $10^{-3}$ | |
|---|---|---|---|---|---|---|---|---|---|---|---|
| 100 (genetic) | -0.187 [-0.24, -0.153] | -0.191 [-0.231, -0.156] | 80.7 [76.2, 84.8] | 89.7 [85.2, 94.2] | 0.539 [0.466, 0.641] | 0.0319 [0.263, 0.388] | 81 [76.6, 84.9] | 7.03 [6.5, 7.52] | 5.97 [5.51, 6.45] | 0.342 [0.271, 0.414] | |
| 1,000 (genetic) | -0.187 [-0.234, -0.153] | -0.182 [-0.227, -0.152] | 80 [75.7, 84.5] | 89.5 [85.3, 93.9] | 0.516 [0.464, 0.561] | 0.317 [0.261, 0.384] | 80.1 [76.6, 84.2] | 7.01 [6.48, 7.49] | 5.97 [5.5, 6.48] | 0.343 [0.274, 0.404] | |
| 100 + 5,000 (combined) | -0.185 [-0.24, -0.147] | -0.19 [-0.253, -0.157] | 82.4 [76.1, 88.6] | 90.5 [85.6, 94.7] | 0.541 [0.458, 0.637] | 0.341 [0.253, 0.461] | 82.7 [76.7, 87.9] | 7.03 [6.5, 7.49] | 6.01 [5.52, 6.49] | 0.362 [0.27, 0.454] | |
| True | -0.2 | -0.2 | 80 | 90 | 5e-05 | 3e-05 | 80 | 7 | 6 | 0.33 | |

# 6. REFERENCES

[1] Yashin I., Arbeev K.G., Wu D., Arbeeva L.S., Kulminski A., Akushevich I., Culminskaya I., Stallard E., Ukraintseva S.V., How lifespan associated genes modulate aging changes: lessons from analysis of longitudinal data, Front. Genet. 2013.

[2] Arbeev, K. G., Ukraintseva, S. V., Kulminski, A. M., Akushevich, I., Arbeeva, L. S., Culminskaya, I. V., et al. (2012). Effect of the APOE polymorphism and age trajectories of physiological variables on mortality: application of genetic stochastic process model of aging. Scientifica 2012:568628. doi: 10.6064/2012/568628.

[3] Deelen, J., Beekman, M., Uh, H.-W., Helmer, Q., Kuningas, M., Christiansen, L., et al. (2011). Genome-wide association study identifies a single major locus contributing to survival into old age; the APOE locus revisited. Aging Cell 10, 686–698.

[4] Breslow N.E and Cain K. C.: Logistic regression for two-stage case-control data, Biometrika (1988) 75 (1): 11-20 doi:10.1093/biomet/75.1.11.

[5] Breslow, N. E. and Holubkov, R. (1997), Weighted likelihood, pseudo-likelihood and maximum likelihood methods for logistic regression analysis of two-stage data. Statist. Med., 16: 103–116. doi:10.1002/(SICI)1097-0258(19970115)16:1<103::AID-SIM474>3.0.CO;2-P

[6] Woodbury, M.A., Manton, K.G.: A random-walk model of human mortality and aging. Theoretical Population Biology 11(1), 37(48), 1977. doi:10.1016/0040-5809(77)90005-3

[7] Fisher L.D. and Lin D.Y.: Time-dependent covariates in the Cox proportional-hazards regression model, Annual Review of Public Health, Vol. 20: 145 -157 (1999).

[8] Yashin, A.I., Arbeev, K.G., Akushevich, I., Kulminski, A., Akushevich, L., Ukraintseva, S.V.: Stochastic model for analysis of longitudinal data on aging and mortality. Mathematical Biosciences 208(2), 538{551 (2007). doi:10.1016/j.mbs.2006.11.006

[9] Yashin, A.I., Arbeev, K.G., Kulminski, A., Akushevich, I., Akushevich, L., Ukraintseva, S.V.: Health decline, aging and mortality: how are they related? Biogerontology 8(3), 291{302 (2007). doi:10.1007/s10522-006-9073-3

[10] Arbeev, K.G., Akushevich, I., Kulminski, A.M., Arbeeva, L.S., Akushevich, L., Ukraintseva, S.V., Culminskaya, I.V., Yashin, A.I.: Genetic model for longitudinal studies of aging, health, and longevity and its potential application to incomplete data. Journal of Theoretical Biology 258(1), 103{111 (2009). doi:10.1016/j.jtbi.2009.01.023

[11] Yashin, A.I., Arbeev, K.G., Akushevich, I., Kulminski, A., Akushevich, L., Ukraintseva, S.V.: Model of hidden heterogeneity in longitudinal data. Theoretical Population Biology 73(1), 1(10) (2008).doi:10.1016/j.tpb.2007.09.001

[12] Witteman, J.C.M., Grobbee, D.E., Valkenburg, H.A., Stijnen, T., Burger, H., Hofman, A., van Hemert, A.M.: J-shaped relation between change in diastolic blood pressure and progression of aortic atherosclerosis. The Lancet 343(8896), 504{507 (1994). doi:10.1016/S0140-6736(94)91459-1. Originally published as Volume 1, Issue 8896

[13] Allison, D.B., Faith, M.S., Heo, M., Kotler, D.P.: Hypothesis concerning the u-shaped relation between body mass index and mortality. American Journal of Epidemiology 146(4), 339{349 (1997)

[14] Boutitie, F., Gueyer, F., Pocock, S., Fagard, R., Boissel, J.P.: J-shaped relationship between blood pressure and mortality in hypertensive patients: New insights from a meta-analysis of individual-patient data. Annals of Internal Medicine 136(6), 438{448 (2002). doi:10.7326/0003-4819-136-6-200203190-00007

[15] Kuzuya, M., Enoki, H., Iwata, M., Hasegawa, J., Hirakawa, Y.: J-shaped relationship between resting pulse rate and all-cause mortality in community-dwelling older people with disabilities. Journal of the American Geriatrics Society 56(2), 367{368 (2008). doi:10.1111/j.1532-5415.2007.01512.x

[16] Mazza, A., Zamboni, S., Rizzato, E., Pessina, A.C., Tikhono, V., Schiavon, L., Casiglia, E.: Serum uric acid shows a j-shaped trend with coronary mortality in non-insulin-dependent diabetic elderly people. the cardiovascular study in the elderly (castel). Acta Diabetologica 44(3), 99{105 (2007). doi:10.1007/s00592-007-0249-3

[17] Okumiya, K., Matsubayashi, K., Wada, T., Fujisawa, M., Osaki, Y., Doi, Y., Yasuda, N., Ozawa, T.: A u-shaped association between home systolic blood pressure and four-year mortality in community-dwelling older men. Journal of the American Geriatrics Society 47(12), 1415{1421 (1999). doi:10.1111/j.1532-5415.1999.tb01559.x

[18] Nelder, John A.; R. Mead (1965). "A simplex method for function minimization". Computer Journal 7: 308–313. doi:10.1093/comjnl/7.4.308.

[19] M. J. D. Powell (2007). A view of algorithms for optimization without derivatives. Cambridge University Technical Report DAMTP 2007.

[20] Akushevich, I., Kulminski, A., Manton, K.G.: Life tables with covariates: Dynamic model for nonlinear analysis of longitudinal data. Mathematical Population Studies 12(2), 51-80 (2005). doi:10.1080/08898480590932296